\documentstyle[11pt,newpasp,twoside,epsf]{article}
\reversemarginpar

\begin{document}
\title{Studies of Low-Mass Star Formation with ALMA}
\author{Neal J. Evans II}
\affil{Department of Astronomy, The University of Texas at Austin
Austin, TX 78712-1083}

\begin{abstract}
ALMA will revolutionize the study of star formation by providing a
combination of angular resolution and sensitivity that far exceeds
that of present instruments. I will focus on studies of relatively
isolated cores that are forming low-mass stars. There is a general
paradigm for the formation of such stars, and there are detailed
theoretical predictions for the evolution of the density and
velocity fields for different assumptions about the initial conditions.
Because the theory is well developed, observational tests are
particularly revealing. The primary probes of physical conditions
in these regions are discussed and the sensitivity of ALMA to these
probes is shown and compared to the current state of the art.
The consequences for the ALMA requirements are discussed.

\end{abstract}

\section{Introduction}

The focus in this paper will be on regions forming low mass stars in
relative isolation, as other papers are covering issues of massive
star formation and clustered star formation. Isolated star formation
is interesting for several reasons.
An empirical evolutionary scheme (Lada 1987, Andr\'e et al. 1993)
is generally accepted.
There is a
well-established theory (Shu et al. 1987) and variations (e.g.,
Foster \& Chevalier 1993; Henriksen, Andr\'e, \& Bontemps 1997; McLaughlin \&
Pudritz 1997) that can be tested by observations.
It is particularly suited to
studies connecting star formation to planet formation.

The two primary probes of the conditions in star-forming cores are continuum
emission from dust and spectral lines from molecules.
These are complementary in many ways. The dust emission is not affected
by molecular depletion and traces column density very effectively, but
dust grain sizes may be a function of the environment, or gas and dust 
distributions may  differ because of ambipolar diffusion.
In principle, molecular spectroscopy probes the local density and
velocity fields, but it is sensitive to variations
in chemical abundances. Together, these two probes can be very  powerful.
With ALMA,
a hybrid probe will become widely available: molecular line absorption
against emission from compact dust components, such as circumstellar
disks. 

The exquisite sensitivity of ALMA to continuum emission will allow us
to map the detailed structure of the dust column density in many cores,
ranging along the evolutionary sequence, to trace the flow of matter from
large scales to disks and stars. Together with information from other
wavelength regions, a complete picture of the distribution of dust
temperature and column density will result, along with information
on possible changes in the grain size distribution.

Maps of optically thin, thermally excited tracers will provide column
densities of gas, for comparison to those of dust measured by the
continuum emission.
The molecular spectroscopy of lines with subthermal excitation will
yield direct estimates of the local density. Combined studies of
optically thick and thin lines will reveal the kinematics in detail.
Finally, absorption spectroscopy of the material in front of opaque
disks will be a new capability of ALMA that will help to unravel the
complex velocity fields involved in forming stars.

The dust continuum emission provides a probe of mass ($M$),
total gas column density ($N$), dust temperature  ($T_D$), grain properties,
and the component of the magnetic field projected on the sky ($B_{\perp}$).
Molecular line emission can probe the gas kinetic temperature ($T_K$),
volume density ($n$), velocity field ($v$), abundances ($X$), and the
line-of-sight 
component of the magnetic field ($B_{\parallel}$). A more detailed
discussion of these probes can be found in Evans (1999).

\section{Dust Continuum Emission}
The dust continuum emission provides a probe of the total amount
and distribution of the dust. If the dust and gas are well mixed,
these quantities can be translated into the same information about the
gas. One caveat is that the dust opacity may change from source to source
or even within a single source (e.g., Visser et al. 1998). 
The primary tool that has been used in
the past is the spectral energy distribution, or SED ($S_{\nu}(\lambda)$),
which gives information on the source luminosity, by integrating
under the SED. The mass can be determined 
by observing the flux in large beams at $\lambda$
large enough that the emission is optically thin. With suitable radiative
transport codes (e.g. Egan, Leung, \& Spagna 1988), source models
constrained by the SED can yield the distribution of the dust temperature
($T_D(r)$) for a given set of grain properties. Matching different parts
of the SED can constrain the choice of grain opacities (Adams, Lada, \& Shu
1987; Butner et al. 1991; van der Tak et al. 1999).

More recently, spatially resolved studies of the intensity of dust
continuum emission have become a
powerful probe. New instruments operating at submillimeter wavelengths
have provided an enormous increase in this kind of data (e.g.,
Johnstone \& Bally 1999; Motte, Andr\'e, \& Neri 1998).
Maps of polarized dust emission are starting to provide
maps of $B_{\perp}$ (Greaves et al. 1999; Rao et al. 1998).
By taking cuts through maps or by azimuthally averaging the intensity, one
obtains $I(b)$, the intensity as a function of impact parameter ($b$),
the separation of the beam from the center of emission.
Plots of $I(b)$ provide probes of column density as a function of
impact parameter $N(b)$, and with modeling, the density distribution
$n(r)$, which is predicted by theories of isolated low-mass star formation.
Recent examples of such studies are those of Chandler \& Richer (2000),
Hogerheijde \& Sandell (2000) and Shirley et al. (2000). They reveal
steeper density gradients than were apparent from most studies of
molecular lines.  An example of
an SED and determinations of $I(b)$ are shown in Fig. 1 for B335,
a well-studied region of low mass star formation. 

\begin{figure}[ht!]
\plotone{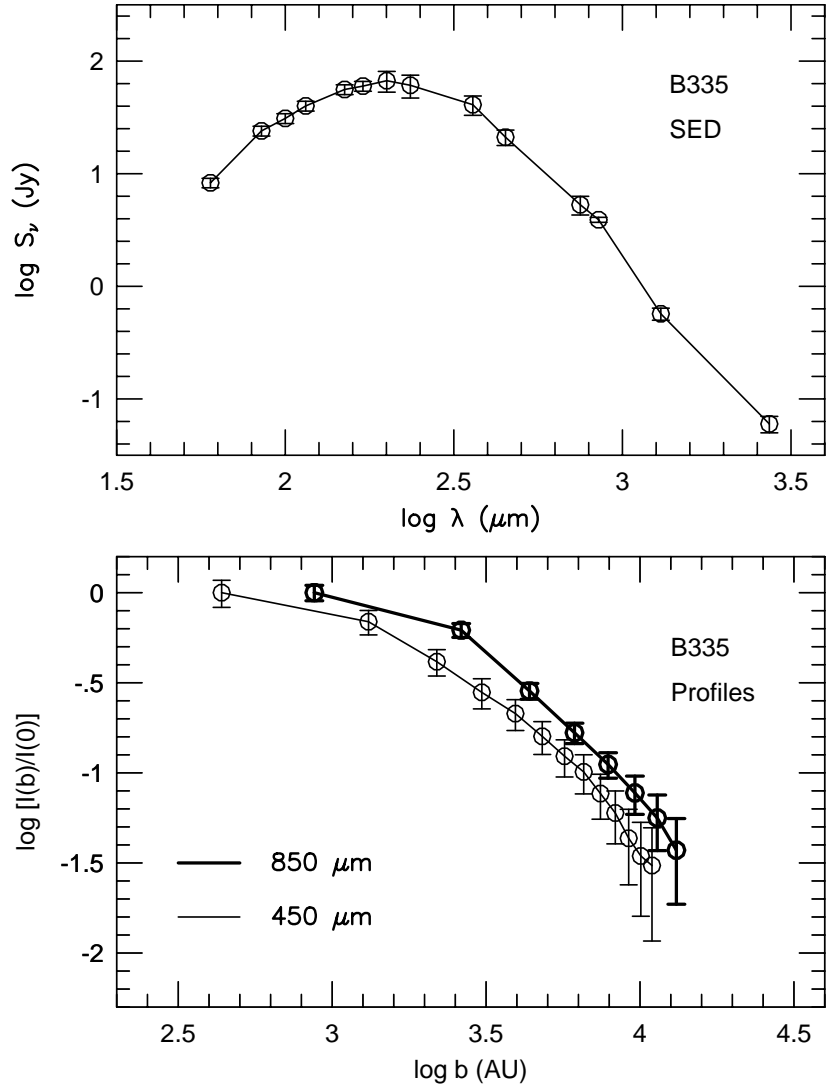}
\caption{The top panel is the SED for B335, an isolated core at a 
distance of 250 pc, which is forming
a low-mass star. References to the photometry can be found in Shirley
et al. (2000). The bottom panel shows the normalized radial intensity
profiles from SCUBA maps at 850 and 450 $\mu$m versus the impact parameter
in AU (Shirley et al. 2000).
}
\end{figure}

Interferometers naturally provide a complementary probe to $I(b)$, the
spatial visibility function ($S(uvdist)$), which can also be compared to
models. This approach is especially well suited to distinguishing
compact structures, such as disks, from the envelope. 
An example of how this capability
can separate components that would be otherwise indistinguishable is given
in Fig. 2, taken from Looney, Mundy, \& Welch (1997).

\begin{figure}[ht!]
%\plotone{looney.ps}
\plotfiddle{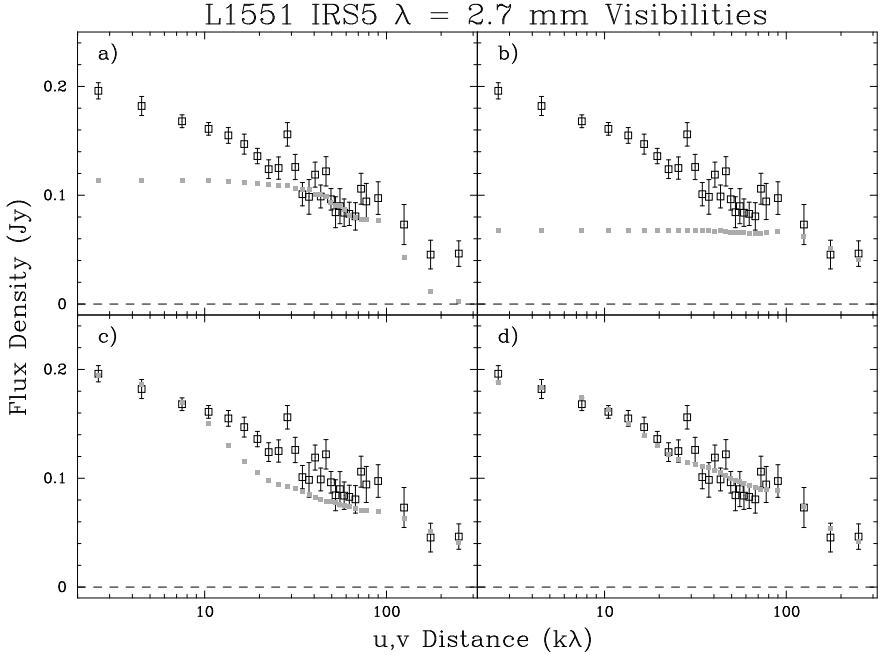}{4.0in}{-90}{50}{50}{-216}{286}
\caption{ Observed visibilities of L1551 IRS 5 at 2.7 mm, binned in annuli
(open
squares with error bars), are plotted versus the projected baseline in units
of $10^3$ times the wavelength. Different models are shown in each panel by
the
small boxes. Panel a has a model with only a Gaussian source of radius 80 AU;
panel b has a model with two point sources constrained to match the map; panel
c adds a truncated power law envelope to the two point
sources; panel d adds to the previous components a circumbinary structure,
represented by a Gaussian. In the final optimization (panel d),
the envelope has a mass of 0.28 M$_{\sun}$ and an outer radius of 1100 AU, the
circumbinary structure has a mass of 0.04 M$_{\sun}$, and the circumstellar
disk masses are 0.024 and 0.009 M$_{\sun}$ (Looney et al 1997).
}
\end{figure}
 
To discuss the sensitivity of ALMA to dust continuum emission, we
make some simplifying assumptions: the Rayleigh-Jeans limit is valid; the
emission is optically thin; the dust opacity follows

\begin{equation}
\kappa_{\nu} = 9.0 \times 10^{-26} \rm{cm}^2 \rm{H}_2^{-1}
\lambda_{mm}^{-1};
\end{equation}
and the resolution is diffraction-limited
($\theta_b = 1.2\lambda/B$, where $B$ is the maximum dimension of the array).
Then the sensitivity to dust emission can be expressed in terms of the
product of gas column density and dust temperature:
 
\begin{equation}
NT_D = 2.5 \times 10^{24} 
\lambda_{\rm{mm}} B^2_{\rm{km}} \Delta S_{\nu}(\rm{mJy}) 
\end{equation}
or, in terms of visual extinction,
 
\begin{equation}
A_VT_D = 2.5 \times 10^{3} 
\lambda_{\rm{mm}} B^2_{\rm{km}} \Delta S_{\nu}(\rm{mJy}).
\end{equation}
Alternatively, one can describe the sensitivity to gas mass in Earth masses
given a distance ($d$):

\begin{equation}
MT_D = 470 M_{\earth} (d/140\rm{pc})^2 
\lambda^3_{\rm{mm}}  \Delta S_{\nu}(\rm{mJy}). 
\end{equation}
The fiducial distance of 140 pc is a typical distance to nearby regions that
are forming low-mass stars.
These equations can be generalized to the case where the Rayleigh-Jeans
approximation fails, at the cost of some clarity. These expressions 
will suffice
to illustrate the main points, if we bear in mind that Rayleigh-Jeans
failure will first tend to decrease the sensitivity to column density or mass
at the shorter wavelengths.

In calculating sensitivities, I have used the values of  
$\Delta S_{\nu}(\rm{mJy})$ given by Butler \& Wootten (1999), for 1.5 mm
of precipitable water vapor (PWV). The values are for 1 $\sigma$
noise after 60 sec of integration. This would correspond to 22 $\sigma$ in
an 8 hour integration. The resulting
plots of $NT_D$ and $MT_D$ versus $\lambda$ are shown in Fig. 3, along with
a plot of the spatial resolution in AU at a distance of 140 pc.
The plot is truncated at 3 mm to allow the shorter wavelengths to be seen 
clearly.
Note that $\lambda \ge 1$ mm have comparable sensitivity to column density,
with much better sensitivity in compact configurations. For the most compact
configuration, ALMA could detect $5.5 \times 10^{20}$ cm$^{-2}$ ($A_V = 0.55$ 
mag) at $T_D = 10$ K.
The sensitivity to mass is best at $\lambda = 0.87$ mm for the standard water
vapor. The mass sensitivity is even better at the shortest wavelengths if
the PWV drops to 0.35 mm (crosses in the top panel), but only 
if the dust is warm enough that the Rayleigh-Jeans limit applies ($T_D >>
41$K at 850 GHz).

\begin{figure}
\plotone{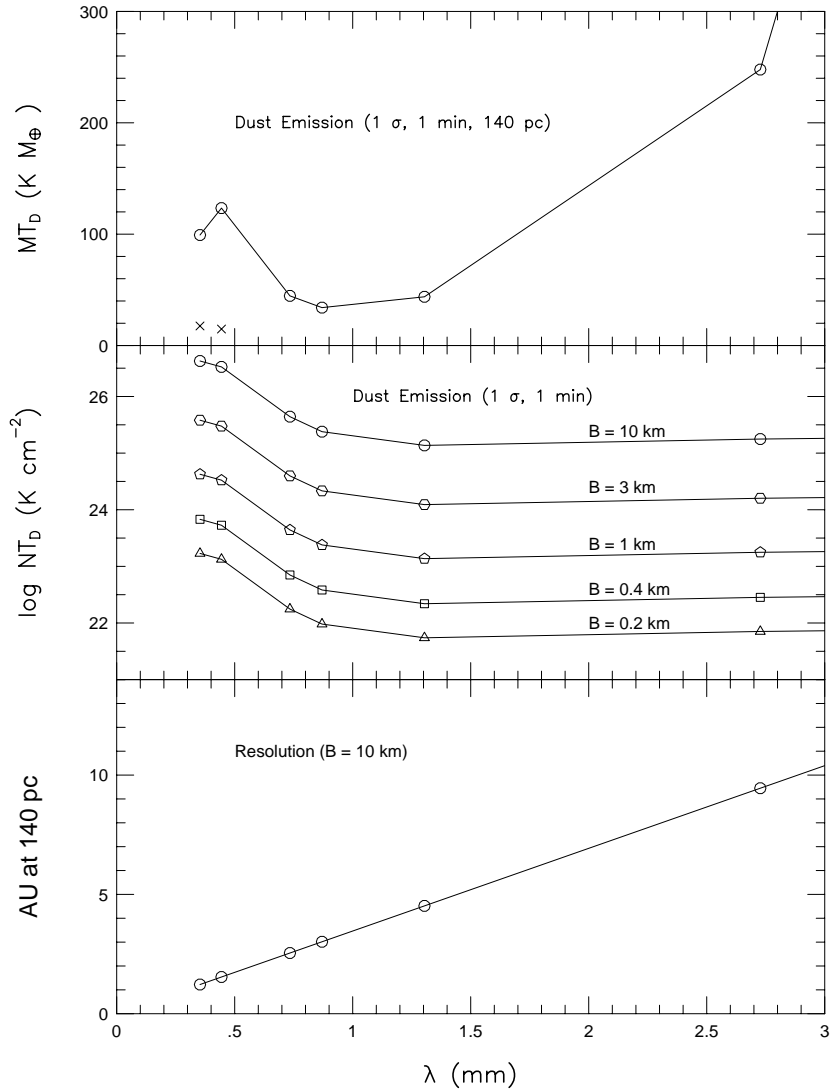}
\caption{Sensitivity to mass (top) versus wavelength for a distance of 140 pc.
The crosses at the bottom left correspond to PWV of 0.35 mm; all other points
are for 1.5 mm.
Sensitivity to column density (middle) versus wavelength, for a maximum 
baseline of 10, 3, 1, 0.4 and 0.2 km. Note the logarithmic scale on this panel.
Bottom plot shows resolution in AU at 140 pc, the 
distance of many star forming regions, for a maximum baseline of 10 km.
}
\end{figure}

To put things in perspective, ALMA can detect the dust emission from 
0.3 $M_{\earth}$ of gas at $T_D = 100$ K at $\lambda = 0.87$ mm! 
Another useful comparison is to the current state of the art.
Table 1 shows the comparison
between ALMA and SCUBA on the JCMT, assuming roughly comparable
PWV, at $\lambda = 0.87$ mm.

\begin{table}
\caption{ALMA versus SCUBA for Dust Continuum Emission }
\begin{tabular}{lrrr}
\tableline
Quantity                     & SCUBA     & ALMA       &    ALMA   \cr
                             &           & $B=0.2$km  & $B=10$km  \cr
\tableline
$\Delta S_{\nu}(mJy)$        & 40        & 0.11       &  0.11     \cr
$\theta_{b}$(arcsec)         & 15        & 1.1        &  0.02     \cr
$\theta_{b}$(AU at 140 pc)   & 2100      & 150        &  3.0      \cr
$A_VT_D$ (mag-K)             & 19        & 9.6        &$2.4\times10^4$  \cr
$M_{\earth} T_D$ (at 140 pc) & $1.2\times 10^4$ & 34  & 34         \cr
\tableline
\end{tabular}
\end{table}

It is clear from the figure and table that ALMA will take us into new
regimes of sensitivity and resolution, allowing study of $N(b)$ and hence
$n(r)$ to much finer scales than currently possible. The key requirement
for this work is spatial dynamic range. Sensitivity to the largest
relevant spatial scales is a challenge for any interferometer, and ALMA
must solve this problem.

\section{Molecular Line Emission}

Mapping molecular line emission provides a wealth of information about
star forming regions. Transitions between levels in thermal equilibrium
(CO, some transitions of H$_2$CO, CH$_3$CN, etc.) provide $T_K$, while
transitions between levels not in thermal equilibrium allow a measure
of $n$. The latter however is coupled to the abundance because of trapping,
and multiple transitions of a single molecule are needed to separate
these effects. In the simplest model, a homogeneous cloud, one derives
a weighted mean density along each line of sight. For more sophisticated
models (e.g., power laws and collapse models), the data constrain the model
parameters.

The line profiles contain vital information about the velocity field.
While this has been difficult to extract, in some cases it is possible
to learn about rotation (Goodman et al. 1993), infall (Myers, Evans, \&
Ohashi 2000), and outflow (Bachiller 1996). For a few transitions,
the line profile, observed with suitable polarization, yields 
$B_{\parallel}$ (e.g., Crutcher 1999).

The key instrumental parameters for molecular line emission
are the beam size and the velocity resolution. Butler \& Wootten
(1999) give the formula for the 
line radiation temperature (proportional to intensity) for the case
of diffraction-limited beams and velocity resolution of 1 km s$^{-1}$:
\begin{equation}
\Delta T_R = 0.32 \rm{K}\ B^2_{\rm{km}} \Delta S_{1}(\rm{mJy}).
\end{equation}

While bright lines may be observed with diffraction-limited beams, choosing
a fixed beam of 1\arcsec\ provides a convenient benchmark for weak lines. 
In this case,
\begin{equation}
\Delta T_R = 0.013 \rm{K}\ {{ \lambda^2_{\rm{mm}} \Delta S_{1}(\rm{mJy})}
\over
{(\theta_b/1\arcsec)^2 \sqrt{\delta v (\rm km\ s^{-1})}}}
\end{equation}
where $\Delta S_{1}(\rm{mJy})$ is the sensitivity (1 $\sigma$ in 60 sec)
in a 1 km s$^{-1}$ band
from Butler \& Wootten (1999), and $\delta v (\rm km\ s^{-1})$ is the
velocity resolution in km s$^{-1}$. 

The current state of the art for interferometers using $\theta_b = 1\arcsec$
is about 1000 mJy at 110 GHz. 
For ALMA, $\Delta S_{1}(\rm{mJy})$ is less than 10 mJy up to 345 GHz, implying
a gain of a factor of 100. Values of $\Delta T_R < 0.5$ K can be achieved
from 0.35 to 2.7mm, even with PWV of 1.5 mm.
Such sensitivity will allow detailed mapping of the temperature, density,
and velocity field at an unprecedented scale.
Figure 4 shows plots of $\Delta T_R$ for $\delta v = 1$ km s$^{-1}$ for
both constant ($\theta_b = 1\arcsec$) and for diffraction-limited beams.

\begin{figure}
\plotone{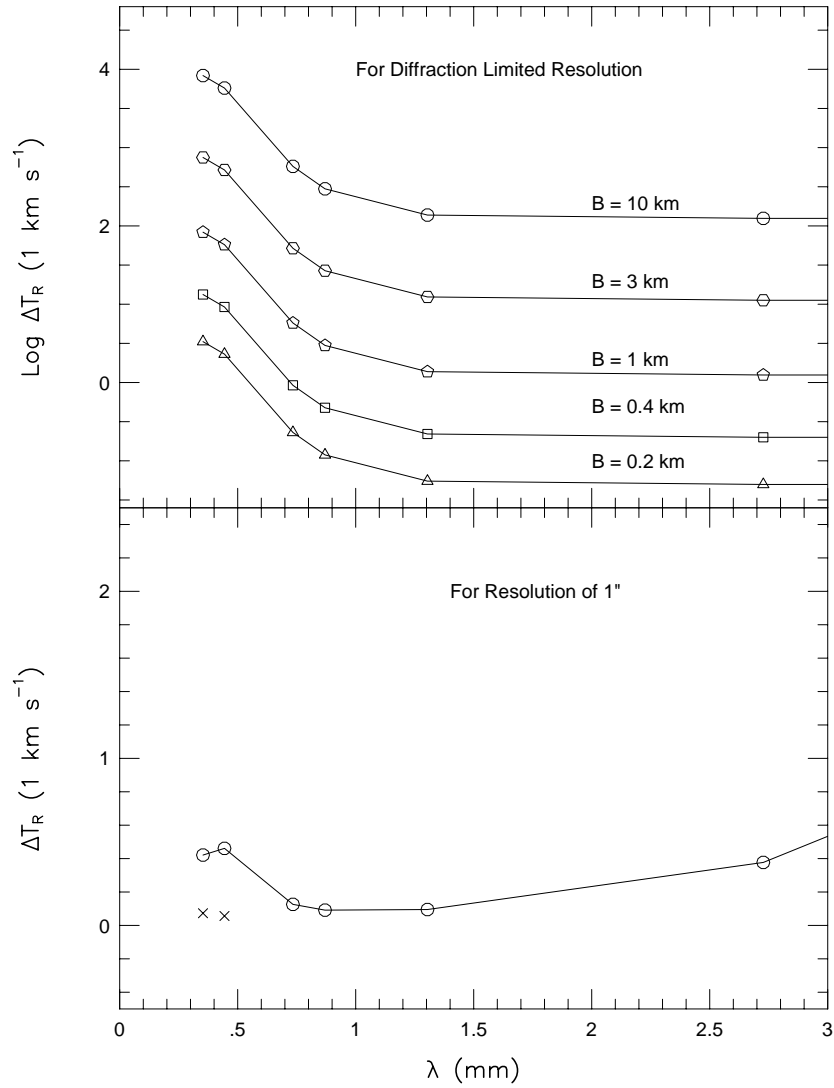}
\caption{Top Panel: sensitivity to line emission versus wavelength for different
maxiumum baselines; note the logarithmic scale for $\Delta T_R$.
Bottom Panel: sensitivity to line emission versus wavelength for 
constant resolution of $\theta_b = 1\arcsec$; note the linear scale
for $\Delta T_R$.
The crosses at the bottom left correspond to PWV of 0.35 mm; all other points
are for 1.5 mm.
All values are for a spectral resolution of 1 km s$^{-1}$.
}
\end{figure}

In addition to the ALMA requirements for sensitivity, spatial resolution, 
and spatial dynamic range established for continuum observations, 
line studies of molecular clouds add velocity resolution (must be easily
variable and very fine) and
frequency coverage (needed to cover the range of transitions and molecules
needed for a full analysis). The baseline ALMA sensitivity for line radiation
observed with diffraction-limited resolution and large baselines 
is marginal. Observing lines
at very high spatial and spectral resolution will stretch ALMA to its
limit in sensitivity.

\section{Molecular Line Absorption Against Continuum Sources}

This technique, familiar at cm wavelengths, has only recently become
possible in regions forming low mass stars (e.g., Choi, Panis, \& Evans 1999). 
The background source 
is a compact dust continuum source, plausibly a circumstellar disk or
perhaps the inner part of the envelope. With current instruments,
only a few disks are strong
enough to produce absorption lines from molecular gas in front.
ALMA will make this a routine probe. Because the disk
lies at the center of the
infalling envelope, only gas in front of the disk 
will show absorption, while the
rest of the cloud will produce emission. This selection provides a 
clear-cut way to resolve the infall-outflow ambiguity that plagues
studies of cloud collapse. For a beam that includes only 
an opaque disk, only the front half of the cloud will be seen in
absorption. For larger beams, the surrounding cloud will produce
emission, resulting in an inverse P-Cygni profile for infall. 
The sensitivities for line emission are sufficient that most disks
can be used in this way.

\section{Summary and Requirements}

ALMA will provide a tremendous advance in capability for studies of
isolated low-mass star formation. The key probes are dust continuum
emission and molecular line emission. A new capability, molecular line
absorption against circumstellar disks, will become routine, allowing
clear-cut resolution of infall-outflow ambiguities.

The requirements on ALMA for dust continuum emission are very low
receiver temperatures and wide bandwidths, coverage of a wide range of
wavelengths, and very good coverage of the uv plane. 
The last of these is particularly
important, as star formation is intrinsically a multiscale problem, with
essential information on scales ranging from a few AU to at least a
few times $10^4$ AU.
For spectral lines,
the large bandwidth requirement is replaced by the need for a flexible
correlator capable of velocity resolution as good as 0.01 km s$^{-1}$.
The requirement for a wide range of wavelengths is stiffened to essentially
complete coverage of the bands that penetrate the atmosphere.

This work has been supported by the State of Texas and NASA grant 
NAG5-7203. L. Looney and Y. Shirley provided figures.
%\par 
%\centerline{\psfig{figure=bossa1.ps,angle=-90,height=3.5in,width=3.5in}}
%\vspace{0.3in}

%\noindent
%manual caption

%\vspace{0.3in}

\end{document}